\documentclass[preprint,aps,showpacs]{revtex4}

\usepackage{graphicx}

\begin{document}
\draft
\title{
Vortex solitons --- Mass, Energy and Angular momentum bunching 
in relativistic electron-positron plasmas
}

\author{T.~Tatsuno, V.~I.~Berezhiani}
\affiliation{Graduate School of Frontier Sciences, The University of Tokyo,
Hongo 7-3-1, Tokyo 113-0033, Japan}

\author{S.~M.~Mahajan}
\affiliation{Institute for Fusion Studies, The University of Texas at Austin,
Austin, Texas 78712}

\date{\today}

\begin{abstract}
It is shown that the interaction of large amplitude electromagnetic waves with a hot electron-positron (e-p) plasma (a principal constituent of the universe in the MeV epoch) leads to a bunching of mass, energy, and angular momentum in stable, long-lived structures.
Electromagnetism in the MeV epoch, then, could provide a possible route for seeding the observed large-scale structure of the universe.
\end{abstract}


\maketitle

\section{Introduction}
It is widely believed that the currently observed large-scale structure of the universe (the clusters and superclusters of galaxies) grew gravitationally out of small density fluctuations \cite{ref-1}.
The imprint of this density variation in early universe was left on the cosmic microwave background radiation in the form of spatial temperature fluctuations.
The gravitational origin, however, can be only a part of the story; the gravity can enhance, but it can not produce these fluctuations.
The quest for the physical process(es) which produced the initial matter-density fluctuations, has led to the emergence of the following two leading mechanisms: inflation, and topological defects \cite{defects,DBM-93}.
Of these the former is, perhaps, the most thoroughly investigated. According to this mechanism there exists, in the evolution of the universe, an early inflationary period in which the universe expands so rapidly (exponentially) that quantum fluctuations get trapped in the expansion.
By the end of the inflation, therefore, small irregularities covering a wide range of length-scales permeate the entire universe.
Gravitational instability then acts on these small initial irregularities and enhances the concentration of matter from which galaxies and clusters of galaxies eventually emerge \cite{PP-93}.
Theory of such processes is, by no means, complete and much needs to be done to determine whether tiny quantum fluctuations can provide a strong enough template for gravitational condensation to finally create the structures that we observe today.
Cosmologist, usually, rely on the speculated existence of non-baryonic dark matter to augment the gravitational force to aid and accelerate the structure formation.
The question is far from settled!

It is natural, then, to look elsewhere for the source of the seed density ``fluctuations.''
An obvious possibility is to explore if electromagnetic interactions taking place in a plasma (known to be the source of a whole variety of linear as well as nonlinear waves) can cause the required density perturbations. In the standard cosmological model of the hot Universe (the Big Bang model), it is estimated that temperatures as high as $T \sim 10^{10} \, {\rm K} \sim 1 \, {\rm MeV}$ prevail up to times $\sim$ 1 second ($t \approx 1 \sec$) after the Big Bang.
In this epoch, the main constituents of the Universe are photons, neutrinos and anti-neutrinos, and e-p pairs \cite{ref-1,PP-93}.
As the plasma cools down the annihilation process $e^+ + e^- \to \gamma + \gamma$ dominates and the $e^+ e^-$ pair concentration goes down.
Since the equilibration rates are fast in comparison with the changes in plasma parameters, an equilibrium e-p plasma should be present in the MeV epoch of the early Universe.
It is this plasma-dominated era in which we will seek the seeds for future structure formation.

Relativistic e-p plasmas have been investigated quite extensively \cite{refs-4}.
Tajima and Taniuti~\cite{TT-90} suggested that collective processes in these plasmas could lead to interesting consequences for structure formation.
In the e-p plasma of the early Universe, localized low frequency electromagnetic (EM) waves ($\hbar \omega \ll T$) could propagate as an envelope soliton due to the interaction with sound waves.
Plasma density variations related to these solitons could potentially be useful towards structure formation in the Universe.
However the analysis \cite{TT-90} is based on a one dimensional formulation, and the corresponding soliton solutions are likely to be unstable in higher dimensions.
Berezhiani and Mahajan~\cite{BM-95} argued that in the MeV epoch of the Universe, although the e-p pairs form the dominant constituent of the plasma, a minority population of heavy ions is also present due to the baryon asymmetry.
They were able to show that, under appropriate conditions (when the plasma is transparent, i.e., $\omega \gg \omega _{e}$, where $\omega$ ($\omega _{e}$) is the pulse (plasma) frequency), the resulting e-p-ion plasma supports the propagation of stable, nondiffracting and nondispersing EM pulses (light bullets) with large density bunching.
It was further shown in Ref.~\cite{MBM-98} that these bullets are exceptionally robust: they can emerge from a large variety of initial field distributions, and are remarkably stable.
Note that the characteristic dimensions of such matter-filled light pulses are proportional to the electron to ion density ratio, and tend to be considerably larger than the skin depth ($\lambda = c / \omega_{e}$).
The implication is that when one deals with the EM structures whose characteristic dimensions (of the spatio-temporal inhomogeneities) are of the order of the skin depth, the baryon asymmetry affects can be safely neglected, and the dynamical system can be assumed as a pure electron-positron plasma.

In the present paper we examine the propagation of strong EM radiation in a hot pure e-p plasma with the explicit aim of finding soliton-type solutions.
The plasma is assumed to be transparent. We demonstrate that the dynamics of the EM field envelope is governed by a generalized nonlinear Schr\"{o}dinger equation (NSE) with a defocusing nonlinearity.
In 1D, this equation admits dark soliton solutions, while in 2D, the so called vortex soliton solutions are also possible.

Dark solitons exist as dips on a continuous-wave background field.
Being stable in one dimensional, they appear as dark-stripe solitary waves in bulk medium.
However, such stripes are unstable to transverse modulations that results in the induced generation of vortices with alternating polarities.

Vortex solitons, which are the most fundamental 2D soliton solutions of NSE with an angular $2 \pi$ phase ramp, appear as local dark minima in an otherwise bright background.
Vortex solitons have been recently observed in materials with a defocusing optical nonlinearity --- the dynamics of laser beams in these materials is generally described by the NSE \cite{KL-98}.
Since the electromagnetic vortices (EMV) carry angular momentum that is conserved during propagation, the generation of Vortex solitons in an e-p plasma is a potent mechanism for creating domains with definite angular momenta even out of an initial field distribution devoid of angular momentum.
To keep the total angular momentum zero, domains of equal and opposite angular momentum must be created in pairs.

\section{Formulation}
We use the following set of relativistic hydrodynamic equations in
dimensionless form ~\cite{BM-95}:
\begin{eqnarray}
	\frac{{\rm d}_{\pm}}{{\rm d}t}(G^{\pm}\gamma^{\pm}) - \frac{1}{n^{\pm}}
	\frac{\partial }{\partial t} P^{\pm} &=& \mp \mbox{\boldmath $v$}^{\pm}
	\cdot \frac{\partial \mbox{\boldmath $A$}}{\partial t} \mp
	(\mbox{\boldmath $v$}^{\pm} \cdot \nabla \phi ),
\label{eq. motion 0} \\
	\frac{{\rm d}_{\pm}}{{\rm d}t}(G^{\pm} \mbox{\boldmath $p$}^{\pm})
	+ \frac{1}{n^{\pm}} \nabla P^{\pm} &=& \mp \frac{\partial
	\mbox{\boldmath $A$}}{\partial t} \pm [ \mbox{\boldmath $v$}^{\pm}
	\times (\nabla \times \mbox{\boldmath $A$}) ] 
	\mp \nabla \phi,
\label{eq. motion 1} \\
	\frac{\partial n^{\pm}}{\partial t}+\nabla \cdot (n^{\pm}
	\mbox{\boldmath $v$}^{\pm}) &=& 0,
\label{eq. continuity}
\end{eqnarray}
along with the field equation (in the Coulomb gauge $\nabla \cdot \mbox{\boldmath $A$} = 0$)
\begin{equation}
	\frac{\partial^{2}}{\partial t^{2}} \mbox{\boldmath $A$}
	- \nabla^{2} \mbox{\boldmath $A$} + \frac{\partial}{\partial t} \nabla
	\phi + ( n^{-} \mbox{\boldmath $v$}^{-} - n^{+} \mbox{\boldmath $v$}^{+} )
	= 0,
\label{eq. Maxwell}
\end{equation}
where $\mbox{\boldmath $p$}^{\pm}=\gamma^{\pm} \mbox{\boldmath $v$}^{\pm}$ with factor $\gamma^{\pm} = [ 1 + (\mbox{\boldmath $p$}^{\pm})^{2}]^{1/2}$, ${\rm d}_{\pm} / {\rm d} t = \partial / \partial t + \mbox{\boldmath $v$}_{\pm} \cdot \nabla$ is the comoving derivative, and $G^{\pm} = K_{3}(1/T^{\pm})/K_{2}(1/T^{\pm})$ with $K_{n}$ the $n\/$-th order modified Bessel functions of the second kind.
The superscript labels the particles, electrons ($-$) and positrons ($+$) respectively.
In these equations the time and space variables are in units of the electron plasma frequency $\omega_{e} = (4 \pi e^{2} n_{0} / m_{e})^{1/2}$ and the collisionless skin depth $c/\omega _{e}$ respectively, the field potentials ($\phi$, $\mbox{\boldmath $A$}$) are in the units of $m_{e}c^{2}/e$, and the relativistic momentum vector $\mbox{\boldmath $p$}^{\pm}$ is in units of $m_{e}c$.
The particle number density $n^{\pm}$ is normalized by the equilibrium density $n_{0} = n_{0}^{\pm}$ and plasma temperature ($T^{\pm}$) is measured in units of $m_{e}c^{2}$.
The pressure $P^{\pm} = n_{\rm r}^{\pm} T^{\pm}$, where $n_{\rm r}^{\pm}$ is the density in the rest frame of the fluid element ($n_{\rm r}^{\pm}=n^{\pm} / \gamma^{\pm}$).
The function $G(z)$ defines the ``effective'' temperature dependent mass of the particles and has the following limiting expressions: $G \approx 1 + 5/2z$ for $z \gg 1$, and $G \approx 4/z$ if $z \ll 1$.

From Eqs.~(\ref{eq. motion 0})-(\ref{eq. continuity}) it is straightforward to derive the adiabatic equation of state \cite{refs-6}: 
\begin{equation}
	\frac{n^{\pm} / T^{\pm}}{\gamma^{\pm} K_{2}(1/T^{\pm})}
	\exp (-G^{\pm} / T^{\pm}) = \mbox{\rm const.},
\label{eq. state}
\end{equation}
which, at nonrelativistic temperature ($T^{\pm} \ll 1$), reduces to the standard adiabatic relation ($n_{\rm r}^{\pm}/(T^{\pm})^{3/2} = {\rm const}$) for an ordinary gas.
In the ultrarelativistic limit ($T^{\pm} \gg 1$), as expected, Eq.~(\ref{eq. state}) describes the photon gas ($n_{\rm r}^{\pm}/(T^{\pm})^{3}={\rm const}$).
In the ultrarelativistic case, one should take into account the radiative pressure $P_{R} = \sigma T^{4}$ ($\sigma = \pi / 45 \hbar^{3} c^{3}$).
For simplicity we neglect this less important effect for the current considerations.
Notice that in the MeV epoch, the plasma temperature $T^{\pm} \approx m_{e} c^{2}$ (i.e. $z \approx 1$), and $G \approx 4$ leading to an effective mass of e-p pairs to be $m_{{\rm eff}} \sim 4m_{e}$.
Since the particle masses are just a few times larger than their rest mass at these temperatures, the e-p plasma can be considered as a two component fluid rather than a photon gas.

We consider the propagation of circularly polarized EM wave with a mean frequency $\omega$, and a mean wave number $k$ along the $z$ axis.
The choice of circular polarization is not restrictive, it simplifies the analysis by preventing harmonic generation.
The vector potential can be represented as:
\begin{equation}
	\mbox{\boldmath $A$}_{\perp} = \frac{1}{2} (\mbox{\boldmath $x$}
	+ {\rm i} \mbox{\boldmath $y$}) \, A_{\perp} (\mbox{\boldmath $r$}_{\perp},
	z,t) \, \exp ({\rm i}kz - {\rm i}\omega t) + {\rm c.c.},
\label{CPEM wave}
\end{equation}
where $A_{\perp}$ is a slowly varying function of $\mbox{\boldmath $r$}$ and $t$ ($k \gg \nabla$, $\omega \gg \partial_{t}$).
The unit vectors $\mbox{\boldmath $x$}$ and $\mbox{\boldmath $y$}$ define two mutually perpendicular axes in the plane normal to the direction of wave propagation.
The Coulomb gauge condition leads to the relation $A_{z} = (i/k) (\nabla_{\perp} \cdot \mbox{\boldmath $A$}_{\perp}) \ll A_{\perp}$.
Consequently the effects related to $A_{z}$ will turn out to be negligibly small.
We shall now follow standard methods to analyze the system.
In the slowly varying amplitude approximation, the transverse, high-frequency component of the equation of motion yields the simple relation between the particle momentum and the vector potential \cite{BM-95},
\begin{equation}
	\mbox{\boldmath $p$}_{\perp}^{\pm} G^{\pm}
	= \mp \mbox{\boldmath $A$}_{\perp}.
\label{curl free}
\end{equation}

The low frequency motion of the plasma is driven by the ponderomotive pressure (${} \sim (\mbox{\boldmath $p$}^{\pm})^{2}$) of the high-frequency EM field, and it does not depend on the sign of the particles' charge.
If we assume that in equilibrium the electron and the positron fluids have equal temperature ($T_{0}^{\pm} = T_{0}$), their effective masses will also be equal ($G^{\pm} = G$), and the radiation pressure will impart equal low-frequency momenta to both fluids allowing the possibility of overall density changes without producing charge separation.
The charge neutrality conditions $n^{-} = n^{+} = N$, $\phi = 0$ will be assumed in the rest of the paper.
It is also evident that the symmetry between the two fluids keeps their temperatures always equal ($T^{\pm}=T$) if they were equal initially.

Considerable simplification results when we invoke the wide beam approximation \cite{refs-7}; we assume that the longitudinal variation of the field envelope is much stronger than the transverse variation, i.e.~$L_{z}$, the characteristic length along the propagation direction, is much shorter than $L_{\bot}$, the characteristic length in the transverse plane.
This approximation, coupled with charge neutrality, allows us to extract from Eqs.~(\ref{eq. motion 0})-(\ref{eq. motion 1}), the following, leading order description for the low frequency response: the equation of motion
\begin{equation}
	\frac{\rm d}{{\rm d}t} Gp + \frac{1}{N} \frac{\partial}{\partial z}
		\frac{NT}{\gamma} = - \frac{1}{2 \gamma G} \frac{\partial
		|A_{\bot}|^{2}}{\partial z},
\label{eq. motion 1-1}
\end{equation}
and the ``energy'' conservation equation: 
\begin{equation}
	\frac{\rm d}{{\rm d}t} G \gamma - \frac{1}{N} \frac{\partial}{\partial t}
		\frac{NT}{\gamma} = \frac{1}{2 \gamma G} \frac{\partial
		|A_{\bot }|^{2}} {\partial t}.
\label{eq. motion 0-1}
\end{equation}
Here we have used the condition that the ponderomotive pressure gives equal longitudinal momenta to both electrons and positrons ($p_{z}^{\pm} = p$).
Notice that the assumed circular polarization of the EM field insures that the relativistic factor $\gamma $ does not depend on the ``fast'' time ($1/\omega$) scale; it can be written as
\begin{equation}
	\gamma = \left[ 1 + \frac{|A_{\bot}|^{2}}{G^{2}} + p^{2} \right]^{1/2}.
\end{equation}

Substituting Eqs.~(\ref{CPEM wave}) and (\ref{curl free}) into Eq.~(\ref{eq. Maxwell}), we find that the slowly varying
amplitude $A_{\bot}$ must satisfy 
\begin{eqnarray}
	&& 2 {\rm i} \omega (\partial_t + v_g \partial_z) A_{\perp} +
		\nabla_{\!\!\perp}^2 A_{\perp} \nonumber \\
	&& {} + (\partial_z^2 - \partial_t^2)
		A_{\perp} + (\omega^2 - k^2) A_{\perp} - \frac{2N}{\gamma G}
		A_{\perp}= 0,  \label{Maxwell 2}
\end{eqnarray}
where $v_g$ denotes the group velocity of the carrier waves, $v_g = k / \omega$.

We are still not quite done with simplifying assumptions.
We seek solutions which vary slowly with time in a frame comoving with the wave, that is, in the frame propagating with the group velocity $v_{g}$.
The transformation $\xi = z - v_{g} t$, $\tau = t$ with the condition $v_{g} \partial_{\xi} \gg \partial_{\tau}$ helps implement this approximation.
Equations (\ref{eq. motion 1-1}) and (\ref{eq. motion 0-1}) can now be combined to derive
\begin{equation}
	\frac{\partial}{\partial \xi}[G(\gamma - v_{g} p)] = 0,
\label{G.par}
\end{equation}
the implied constant of motion is to be determined from the boundary conditions.
We demand $p$ and $A_{\perp}$ to be zero at infinite $\xi$, but allow them to be finite as $r_{\perp} \to \infty$.
Integrating Eq.~(\ref{G.par}) leads to ($T_{0}$ is the particle temperature at infinity)
\begin{equation}
	G(T)(\gamma - v_{g} p) = G_{0} (T_{0}),
\end{equation}
which is readily solved for an explicit expression for the longitudinal momentum in terms of the transverse vector potential
\begin{equation}
	p = v_{g} \gamma _{g}^{2} \frac{G_{0}}{G} \biggl[ 1 - \frac{1}{G_{0}v_{g}
	\gamma _{g}}(\gamma_{g}^{2} G_{0}^{2} - G^{2} - |A_{\bot }|^{2})^{1/2}
	\biggr],
\label{momentum}
\end{equation}
where $\gamma_{g} = 1 / (1-v_{g}^{2})^{1/2}$ is the ``effective relativistic factor'' associated with the group velocity of the wave; it is not to be confused with the particle $\gamma$.
The continuity equation can be similarly integrated to determine the particle density (after using Eq.~(\ref{momentum}) for $p$): 
\begin{equation}
	\frac{N}{\gamma G} = \frac{v_{g}\gamma_{g}}{(\gamma_{g}^{2} G_{0}^{2}
	- G^{2} - |A_{\perp}|^{2})^{1/2}}.
\label{plasma current}
\end{equation}

Substituting Eq.~(\ref{plasma current}) into Eq.~(\ref{Maxwell 2}), we obtain the following nonlinear Schr\"{o}dinger equation for the complex amplitude $A_{\perp}$,
\begin{eqnarray}
	&& 2 {\rm i} \omega \partial_{\tau} A_{\perp} + \nabla_{\!\!\perp}^{2}
	A_{\perp} + \frac{1}{\gamma_{g}^{2}} \partial_{\xi}^{2} A_{\perp}
	\nonumber \\
	&& {} + \frac{2}{G_{0}} \biggl( 1 - \frac{v_{g} \gamma_{g}
	G_{0}}{(\gamma_{g}^{2} G_{0}^{2} - G^{2} - |A_{\perp }|^{2})^{1/2}}
	\biggr) A_{\perp} = 0,
\label{NSE}
\end{eqnarray}
where the wave frequency $\omega$ satisfies the dispersion relation $\omega^{2}=k^{2} + {2}/{G_{0}}$, implying that the parameter $\gamma_{g} = \omega \sqrt{ G_{0}/2}$ ($\gamma_{g} = (\omega / \omega _{e}) \sqrt{G_{0}/2}$ in physical quantities).
The set of Eq.~(\ref{NSE}), and the equation of state (\ref{eq. state}) (in which the relation (\ref{plasma current}) could be easily incorporated) constitutes a complete description of the dynamics of strong EM waves in relativistic e-p plasma in the wide beam approximation.

We remind the reader that Eq.~(\ref{NSE}) was derived under the assumption $\partial_{\xi} \gg \nabla_{\!\!\perp}$ (i.e.~$L_{z} \ll L_{\bot})$.
In spite of that, for a highly transparent plasma ($\gamma_{g} \gg 1$), the second, ``diffractive'' term can be the same order or even greater than the third, ``dispersive'' term.
For this paper, we will not attempt the general solutions of this quite complicated set of equations; we will simply deal with waves for which the plasma is so highly transparent that the diffractive term dominates.
Using $\gamma_{g} G_{0} \gg G$, and neglecting the dispersive term, the NSE simplifies to
\begin{equation}
	{\rm i} \partial_{\tau} A_{\perp} + \frac{1}{2} \nabla_{\!\!\perp}^2
	A_{\perp} - 2 \left[ \left( 1 - \frac{|A_{\perp}|^2}{
	\gamma_g^2 G_0^2} \right)^{-1/2} - 1 \right] A_{\perp} = 0,
\label{2D-NSE}
\end{equation}
where the following renormalizations are used: $\tau / 2 \omega G_0 \to \tau$, $\mbox{\boldmath $r$}_{\bot} / \sqrt{2 G_0} \to \mbox{\boldmath $r$}_{\bot}$.

The vector potential $|A_{\perp }|$ is restricted from above by the condition $|A_{\perp}| < \gamma_{g} G_{0}$.
This restriction is necessary for the validity of the hydrodynamic treatment for the particles.
For larger amplitudes, the electromagnetic waves are overturned causing multistream motion of the plasma requiring a kinetic description.
Notice that despite the upper bound on the amplitude of the vector potential, the EM field can be still relativistically strong, i.e., the normalized $|A_{\perp }| \gg 1$, since $\gamma_{g} \gg 1$.

\section{Stationary Solutions}
\label{solutions}
In the NSE derived above the diffractive and nonlinear terms have opposite signs and as a consequence Eq.~(\ref{2D-NSE}) does not admit transversely localized solutions (also called the bright solitons).
Any localized initial EM field, therefore, will undergo transverse spreading during propagation.
The NSE with a defocusing nonlinearity can, however, support stationary structures with asymptotically (at infinity) nonvanishing fields.
Dark solitons in 1D, and vortex solitons in 2D, are the fundamental representatives of such solutions.
In the extreme low amplitude limit, $|A_{\perp}| \ll \gamma_{g} G_{0}$, Eq.~(\ref{2D-NSE}) reduces to the NSE with a cubic nonlinearity.
In one-dimensional geometry we have: 
\begin{equation}
	{\rm i} \frac{\partial A_{\perp}}{\partial \tau}
	+ \frac{1}{2} \frac{\partial^{2} A_{\perp}}{\partial x^{2}}
	- \frac{1}{\gamma_{g}^{2} G_{0}^{2}} |A_{\perp}|^{2} A_{\perp } = 0,
\label{1D-NSE}
\end{equation}
This equation is exactly integrable via the inverse scattering method~\cite{ZS-73}, and its one-soliton solution can be written as~\cite{KL-98}
\begin{equation}
	A_{\perp}(x,\tau) = \gamma_{g} G_{0} A_{0}
	(\alpha \tanh \Theta + {\rm i} \beta) \,
	e^{-{\rm i} A_{0}^{2} \tau},
\end{equation}
where 
\begin{equation}
	\Theta = \alpha A_{0}(x - \beta A_{0} \tau).
\end{equation}
Here $A_{0}$ is a measure of the asymptotic fields at the spatial infinity and $\alpha$ and $\beta$ are constants with $\alpha^{2} + \beta^{2} = 1$.
The solutions with a nonzero value at the center of the dip, is termed the `gray soliton' to distinguish it from the `black soliton' (zero amplitude at the dip) corresponding to $\beta = 0$.
The dark solitons of this class of NSE's do not have any threshold values for their excitation unlike bright solitons (of the appropriate equations) which do.
In other words, dark solitons can be created by an arbitrary small initial dip on a homogeneous background.

In two transverse dimensions (2D), a dark soliton represents a dark stripe imposed on a homogeneous bright background.
It is well-known that such a stripe is unstable to transverse, long wave length modulations \cite{transverse}.
The instability causes the stripe to split into a sequence of Vortex solitons of alternative polarities.
The vortices are dark holes on a bright background, with a nested phase dislocation of the order $m = \pm 1, \pm 2, \ldots$ at their core.

The Vortex soliton solutions of NSE were first suggested by Pitaevskii \cite{LP-61} as topological excitations in an imperfect Bose gas in the superfluids.
The ability of some electromagnetic systems (like the e-p plasma) to simulate fluid dynamical phenomena (like vortex formation) can be demonstrated by applying the Madelung transformation: $A_{\bot} = \sqrt{\rho} \exp ({\rm i} \psi)$ to the defining equations \cite{refs-10}.
The transformation converts the original set to one that is similar to the fluid hydrodynamic equations with a fluid ``density'' $\rho$, and fluid ``velocity'' $\mbox{\boldmath $v$} = \nabla_{\!\!\perp} \psi$.
Vortices can exist despite the potential nature of the ``fluid'' flow.
Indeed, the Madelung transform is singular at the points where $\rho = 0$; these are just branch points where the real and the imaginary parts of the field become zero, while the velocity circulation $\oint \mbox{\boldmath $v$} \cdot \, {\rm d} \mbox{\boldmath $l$} = 2 \pi m$, where the integration is done on a closed path enclosing the singular point, and the integer $m$ is known as the topological ``charge'' of vortex.
Thus the vortex soliton is a topological structure; it can disappear only when annihilated by a vortex soliton of the opposite ``charge.''
The development of the transverse instability of a dark soliton has close parallels in hydrodynamics: for instance the Kelvin-Helmholtz instability, which occurs when the boundary between two flows develops the so-called vortex streets \cite{refs-10}.
Since a dark solitary stripe does not carry any topological ``charge,'' it is evident that vortices have to be born with equal and opposite topological ``charges.''

It is straightforward to show that e-p plasmas can support large amplitude dark solitons as well.
In the general case (amplitude large, but subject to the condition $|A_{\perp}| < \gamma_{g} G_{0}$), we cannot construct analytic solutions even in 1D.
It is possible, however, to extract the general properties of the solution by using reasonably simple techniques especially when $A_{\perp}$ has the time dependence
\begin{equation}
	A(x,\tau) = \hat{A}(x) \exp (- {\rm i} \lambda \tau),
\label{decompose}
\end{equation}
where $A = A_{\perp} / \gamma_{g} G_{0}$ is the normalized amplitude, and is always less than unity.
Here $\lambda$ is so-called nonlinear frequency shift.
This time-dependence implies that the amplitude square is stationary (what follows, therefore, are classed as stationary solutions), and the dip of the wave does not propagate in the comoving frame with quite the group velocity of the linear wave.

\begin{figure}
\begin{center}
\includegraphics[width=12cm]{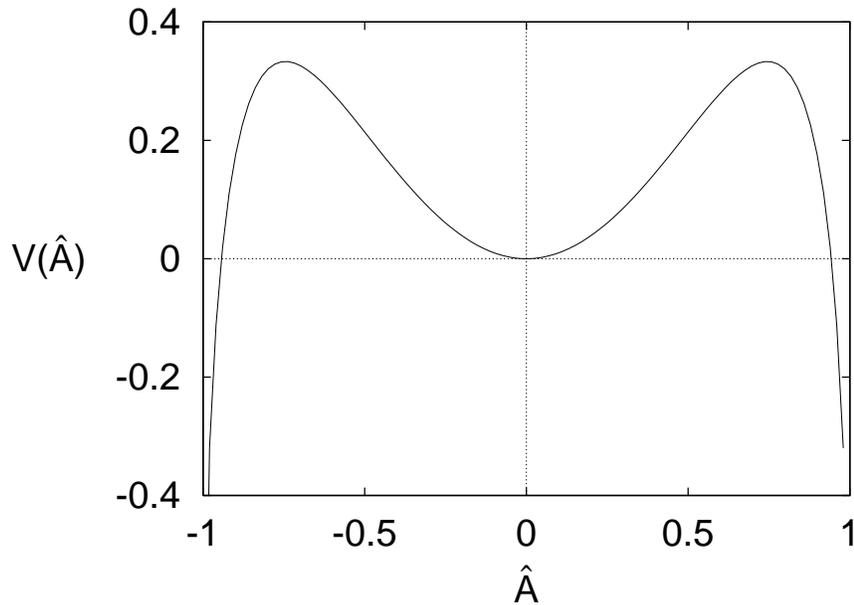}
\caption{The form of the potential $V(\hat{A})$ in Eq.~(\ref{1D-Newtonian}) is shown for $\lambda = 1$.}
\end{center}
\label{fig.1}
\end{figure}

The 1D Eq.~(\ref{1D-NSE}) now can be cast in the form 
\begin{equation}
	\frac{{\rm d}^{2}}{{\rm d} x^{2}} \hat{A} + V'(\hat{A}) = 0,
\label{1D-Newtonian}
\end{equation}
where the prime on $V$ denotes the derivative with respect to $\hat{A}$, and 
\begin{equation}
	V(\hat{A}) = ( \lambda + 2 ) \hat{A}^{2} + 4 \sqrt{1 - \hat{A}^{2}} - 4,
\label{potential representation}
\end{equation}
denotes the potential.
The resemblance of Eq.~(\ref{1D-Newtonian}) to the one obeyed by a Newtonian particle in a nonlinear potential suggests an obvious method for analysis.
One can easily prove that bounded solution exists provided nonlinear frequency shift is positive ($\lambda > 0$).
The profile of the potential, shown in Fig.~\ref{fig.1} for $\lambda = 1$, reveals that the dark soliton solution may reside in the potential well.
Equating ${\rm d}^{2} \hat{A} / {\rm d} x^{2}$ to zero, we estimate the upper bound on $\hat{A}$
\begin{equation}
	\hat{A}_{\rm ub} = \sqrt{1 - \biggl( \frac{2}{\lambda + 2} \biggr)^{2}}.
\label{upper bound}
\end{equation}
The lowermost value of the amplitude $|A|$ is zero, that is, we recover the dark soliton.
Note that for small values of $\lambda$ (${} \ll 1$), $\hat{A}_{\rm ub} \to 0$, while $\hat{A}_{\rm ub} \to 1$ for $\lambda \gg 1$.

\begin{figure}
\begin{center}
\includegraphics[width=12cm]{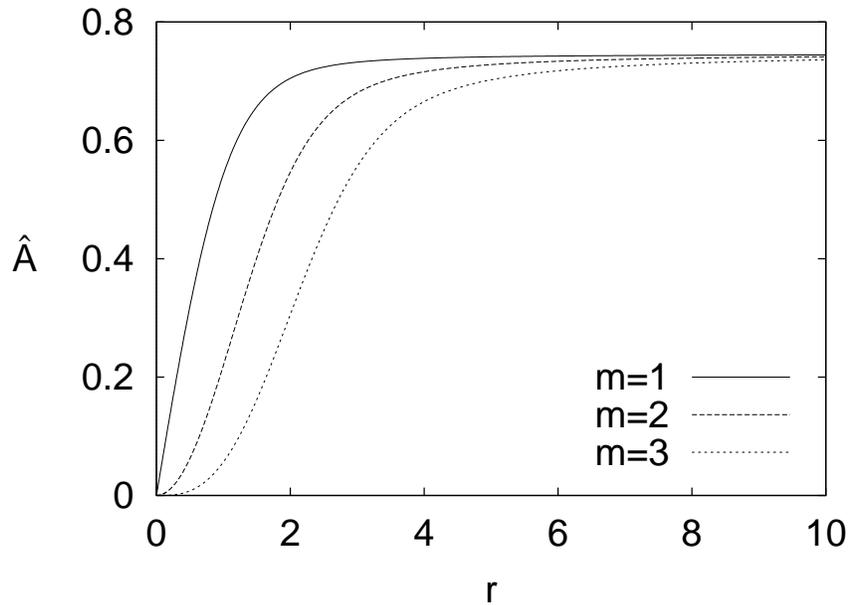}
\caption{The form of the soliton-like solutions for the normalized potential $\hat{A}$ is shown for $m = 1$, $2$, $3$ in case of $\lambda = 1$.}
\end{center}
\label{fig.2}
\end{figure}

Similarly one can show the existence of vortex soliton solutions in 2D.
We shall again seek stationary solutions in 2D polar coordinates $(r,\theta)$.
The ansatz
\begin{equation}
	A = \hat{A}(r) \exp ( {\rm i} m \theta - {\rm i} \lambda \tau ),
\end{equation}
with $\hat{A}(r)$ real, and with the perpendicular Laplacian operator given by 
\begin{equation}
	\nabla_{\!\!\perp}^{2} = \frac{{\rm d}^2}{{\rm d}r^2} + \frac{1}{r}
	\frac{\rm d}{{\rm d}r} - \frac{m^{2}}{r^{2}},
\end{equation}
converts Eq.~(\ref{2D-NSE}) to the ordinary differential equation (similar to Eq.~(\ref{1D-Newtonian})),
\begin{equation}
	\frac{{\rm d}^2}{{\rm d}r^2} \hat{A} + V'(\hat{A}) = - \frac{1}{r}
	\frac{{\rm d} \hat{A}} {{\rm d}r} + \frac{m^{2}}{r^{2}} \hat{A},
\label{2D-Newtonian}
\end{equation}
where the potential $V(\hat{A})$ is the same as the one-dimensional expression given by Eq.~(\ref{potential representation}).
If we were to extend the ``particle in a potential'' analogy further, Eq.~(\ref{2D-Newtonian}) could be viewed as the nonconservative motion of a particle.
Since the rhs approaches zero in the limit $r \to \infty$, Eq.~(\ref{2D-Newtonian}) gives precisely the 1D asymptotic value (Eq.~(\ref{upper bound}) for the vector potential $\hat{A}$).
The behavior at the origin ($r=0$) is totally different; the regular singular point at the origin $r=0$ forces the acceptable $\hat{A}$ to vanish for $m \ge 1$ as $r^{m}$.
The numerical solutions of the two-dimensional nonlinear Schr\"{o}dinger equation for $m=1$, $2$, $3$ are shown in Fig.~\ref{fig.2}.
As expected, the soliton-like solutions evidently go to zero as $r^{m}$ for small $r$, and reach an $m$-independent asymptotic value predicted by Eq.~(\ref{upper bound}).

More general aspects of the dynamics of the EM field can be studied mainly through numerical simulations of Eq.~(\ref{2D-NSE}) --- this is beyond the intended scope of this paper.
We would content ourselves here by making a few qualitative remarks and pointing out directions for the future efforts.
The nonlinearity in Eq.~(\ref{2D-NSE}) is a faster growing function (faster than the cubic) of the field amplitude, and does not exhibit saturation.
We would predict then, that the development of the vortex chain structures from the dark stripe soliton instability for the general system will be faster in comparison with the low amplitude (cubic nonlinearity) case.
Stability of vortex soliton solution is another issue that has to be dealt with.
It is usually believed, though, that vortices with $m = \pm 1$ are topologically stable whereas vortices with larger value of the ``charge'' $m$ may decay into
``single-charge'' vortices.

In three dimensions (3D) the vortices form the so-called vortex line (it looks like a pancake with a hole in the center hanging and moving along the wire-vortex line).
Effects related to the finite group velocity dispersion may lead to a transverse instability of the vortex line.
All these interesting effects are left for future studies.

We would like to emphasize that the electromagnetic fields associated with the dark and vortex solitons are asymptotically (at infinity) nonvanishing.
Due to the generally accepted requirement that in physical system, the fields be localized in all directions, these objects have received much less attention than their localized cousins.
However in recent experiments studying the laser field dynamics in different kinds of optical media, it has been demonstrated that dark and vortex solitons can be readily created as superimpositions upon a localized field background \cite{KL-98}.
This background can be just a few times wider than the soliton width.
During the propagation the background spreads out, reducing its own intensity.
In the light of these experiments let us try to put in perspective the current study of the dark and vortex solitons in e-p plasmas in the early Universe.
Because the typical scale length of these solitons is the collisionless skin depth, we would need a supporting background spanning several skin depths.
This should pose no problem because the ambient uniform field background could easily foot the bill.
The next scale length on which we encounter ``bullet-like'' electromagnetic structures (which owe their origin to the baryon asymmetry) is considerably larger than the skin depth.
Thus the dark and vortex solitons can propagate in a slowly changing background (spreading and decreasing in intensity with the diffractive spreading rate of the soliton decreasing as the background expands), adiabatically maintaining their properties, till they hit the baryon-asymmetry scalelengths.

Topological considerations will insure the preservation of the singular points during propagation.
In the propagating vortex chains, the vortices can move away from one another reducing the possibility of their mutual annihilation.
The propagation introduces elements similar to Hubble expansion --- the structures run away from one another.
These highly speculative remarks need careful investigation.
It is possible that the spreading of the background field may just affect the Vortex distribution and only the cosmological expansion will drive them apart.

What is extremely significant is that during the evolution of the fields, the integrals of motion should be preserved.
It is easy to prove, by direct calculations, that Eq.~(\ref{2D-NSE}) conserves the angular momentum $\mbox{\boldmath $M$}$
\begin{equation}
	(\mbox{\boldmath $M$})_z = \frac{\rm i}{2} \int {\rm d}
	\mbox{\boldmath $r$}_{\bot} \, [\mbox{\boldmath $r$}_{\bot} \times
	(A_{\bot}^* \nabla_{\!\!\perp} A_{\bot} - {\rm c.c.})]_{z}.
\label{angular momentum}
\end{equation}
Equation (\ref{angular momentum}) for the angular momentum is the paraxial approximation for the orbital angular momentum, $\mbox{\boldmath $M$}_{\!\! E} = \int {\rm d} \mbox{\boldmath $r$} \, [ \mbox{\boldmath $r$} \times (\mbox{\boldmath $E$} \times \mbox{\boldmath $B$}) ]$, of the EM field \cite{AB-92}.
The angular momentum carried by the vortices is $M_z = mN$, where $N$ is another conserved quantity known as the ``photon number'' $N = \int {\rm d} \mbox{\boldmath $r$}_{\bot} \, |A_{\bot}|^{2}$ \cite{conv}.

It follows, then, that the relativistic e-p plasma are capable of sustaining electromagnetic vortex like structures, and that these structures have domains in which the EM fields carry non-zero angular momenta, although the total angular momentum of the entire system is zero.
If this angular momentum could, somehow, be locally transferred to the surrounding medium, we would have a rather effective mechanism of imparting angular momentum to different domains of matter in the early universe.
In our next publication we show that when the baryon asymmetry effects are incorporated, the medium can, indeed, acquire angular momentum from the EM field vortices.

\section{Conclusions}
\label{summary}
We have investigated the dynamics of the highly relativistic ($\gamma_{g} \gg 1$) nonlinear propagation of electromagnetic waves in unmagnetized hot electron-positron plasmas.
The system is described by a nonlinear Schr\"{o}dinger equation (\ref{2D-NSE}) with an inverse square root type (non-saturating) nonlinearity.
We have shown the possibility of dark and vortex soliton type solutions for this equation.
Transverse instability of dark soliton stripes leads to the formation of a vortex chain such that the EM fields in each vortex carry angular momentum.
Such objects could play an important role in cosmology as sources for the structure formation in the MeV epoch of the evolution of the Universe.
In the commonly adopted cosmological scenarios about the origin of the rotation of galaxies, structures grow in a hierarchy by the gravitational assembly of clumps out of subclumps.
The origin of the angular momentum of galaxies if they were formed from initial fluctuations in a Friedman Universe was suggested (by Hoyle \cite{FH-49}) to be due to the tidal interactions between the condensing system \cite{PP-93}.
However, it is still not clear whether this mechanism gives an adequate solution \cite{HBN-00}.
We hope that the suggested mechanism of angular momentum generation in the MeV epoch of the Universe is an interesting alternative to explore and examine.
Electromagnetism, operating through the versatile substrate of the e-p plasma, seems to readily generate these highly interesting, long-lived objects --- the carriers of large amounts of mass, energy and angular momentum.
Since an initial localization of mass, energy and angular momentum is precisely the seed that gravity needs for eventual structure-formation, electromagnetism may have provided a key element in the construction of the large-scale map of the observable universe.

Results of this paper can be also applied to astrophysical objects like the pulsars, and active galactic nuclei --- the e-p pairs are thought to be a major constituent of the plasma emanating both from the pulsars, and from the inner region of the accretion disks surrounding the central black holes.


\end{document}